\documentclass[prl,reprint,superscriptaddress,showpacs]{revtex4-1}

\usepackage{graphicx,amsmath,amssymb,bm}
\usepackage{subfigure}
\usepackage{dcolumn}
\usepackage{hyperref}

\newcolumntype{.}{D{.}{.}{1.13}}
\newcolumntype{,}{D{.}{.}{5.2}}

\label{} 

\begin{document}

\title{New precision mass measurements of neutron-rich calcium and potassium
isotopes and three-nucleon forces}

\author{A.T.~Gallant}
\email[Corresponding author:~]{agallant@triumf.ca}
\affiliation{TRIUMF, 4004 Wesbrook Mall, Vancouver, British Columbia, V6T 2A3 Canada}
\affiliation{Department of Physics and Astronomy, University of British Columbia, Vancouver, British Columbia, V6T 1Z1 Canada}
\author{J.C.~Bale}
\affiliation{TRIUMF, 4004 Wesbrook Mall, Vancouver, British Columbia, V6T 2A3 Canada}
\affiliation{Department of Chemistry, Simon Fraser University, Burnaby, BC V5A 1S6, Canada}
\author{T.~Brunner}
\affiliation{TRIUMF, 4004 Wesbrook Mall, Vancouver, British Columbia, V6T 2A3 Canada}
\author{U.~Chowdhury}
\affiliation{TRIUMF, 4004 Wesbrook Mall, Vancouver, British Columbia, V6T 2A3 Canada}
\affiliation{Department of Physics and Astronomy, University of Manitoba, Winnipeg, Manitoba, R3T 2N2 Canada}
\author{S.~Ettenauer}
\affiliation{TRIUMF, 4004 Wesbrook Mall, Vancouver, British Columbia, V6T 2A3 Canada}
\affiliation{Department of Physics and Astronomy, University of British Columbia, Vancouver, British Columbia, V6T 1Z1 Canada}
\author{A.~Lennarz}
\affiliation{TRIUMF, 4004 Wesbrook Mall, Vancouver, British Columbia, V6T 2A3 Canada}
\affiliation{Institut f\"{u}r Kernphysik Westf\"{a}lische Wilhelms-Universit\"{a}t, 48149 M\"{u}nster, Germany}
\author{D.~Robertson}
\affiliation{TRIUMF, 4004 Wesbrook Mall, Vancouver, British Columbia, V6T 2A3 Canada}
\author{V.V.~Simon}
\affiliation{TRIUMF, 4004 Wesbrook Mall, Vancouver, British Columbia, V6T 2A3 Canada}
\affiliation{Fakul\"{a}t f\"{u}r Physik und Astronomie, Ruprecht-Karls-Universit\"{a}t Heidelberg, 69120 Heidelberg, Germany}
\affiliation{Max-Planck-Institut f\"{u}r Kernphysik, 69117 Heidelberg, Germany}
\author{A.~Chaudhuri}
\affiliation{TRIUMF, 4004 Wesbrook Mall, Vancouver, British Columbia, V6T 2A3 Canada}
\author{J.D.~Holt}
\affiliation{Department of Physics and Astronomy, University of Tennessee,
Knoxville, TN 37996, USA}
\affiliation{Physics Division, Oak Ridge National Laboratory, P.O.~Box 2008,
Oak Ridge, TN 37831, USA}
\author{A.A.~Kwiatkowski}
\affiliation{TRIUMF, 4004 Wesbrook Mall, Vancouver, British Columbia, V6T 2A3 Canada}
\author{E.~Man\'{e}}
\affiliation{TRIUMF, 4004 Wesbrook Mall, Vancouver, British Columbia, V6T 2A3 Canada}
\author{J.~Men\'{e}ndez}
\affiliation{Institut f\"ur Kernphysik, Technische Universit\"at
Darmstadt, 64289 Darmstadt, Germany}
\affiliation{ExtreMe Matter Institute EMMI, GSI Helmholtzzentrum f\"ur
Schwerionenforschung GmbH, 64291 Darmstadt, Germany}
\author{B.E.~Schultz}
\affiliation{TRIUMF, 4004 Wesbrook Mall, Vancouver, British Columbia, V6T 2A3 Canada}
\author{M.C.~Simon}
\affiliation{TRIUMF, 4004 Wesbrook Mall, Vancouver, British Columbia, V6T 2A3 Canada}
\author{C.~Andreoiu}
\affiliation{Department of Chemistry, Simon Fraser University, Burnaby, BC V5A 1S6, Canada}
\author{P.~Delheij}
\affiliation{TRIUMF, 4004 Wesbrook Mall, Vancouver, British Columbia, V6T 2A3 Canada}
\author{M.R.~Pearson}
\affiliation{TRIUMF, 4004 Wesbrook Mall, Vancouver, British Columbia, V6T 2A3 Canada}
\author{H.~Savajols}
\affiliation{GANIL, Boulevard Henri Becquerel, Bo\^{i}te Postale 55027, F-14076 Caen Cedex 05, France}
\author{A.~Schwenk}
\affiliation{ExtreMe Matter Institute EMMI, GSI Helmholtzzentrum f\"ur
Schwerionenforschung GmbH, 64291 Darmstadt, Germany}
\affiliation{Institut f\"ur Kernphysik, Technische Universit\"at
Darmstadt, 64289 Darmstadt, Germany}
\author{J.~Dilling}
\affiliation{TRIUMF, 4004 Wesbrook Mall, Vancouver, British Columbia, V6T 2A3 Canada}
\affiliation{Department of Physics and Astronomy, University of British Columbia, Vancouver, British Columbia, V6T 1Z1 Canada}

\date{\today} 

\begin{abstract}
We present precision Penning-trap mass measurements of neutron-rich
calcium and potassium isotopes in the vicinity of neutron number
$N=32$. Using the TITAN system the mass of 
$^{51}$K was measured for the first time, and the precision of the 
$^{51,52}$Ca mass values were improved significantly.
The new mass values show a dramatic increase of the binding energy
compared to those reported in the atomic mass evaluation. In
particular, $^{52}$Ca is more bound by 1.74~MeV, and the behavior
with neutron number deviates substantially from the tabulated values.
An increased binding was predicted recently based on calculations
that include three-nucleon (3N) forces. We present a comparison to
improved calculations, which agree remarkably with the evolution of
masses with neutron number, making neutron-rich calcium isotopes
an exciting region to probe 3N forces at neutron-rich extremes.
\end{abstract}

\maketitle

The neutron-rich calcium isotopes present a key region for
understanding shell structures and their evolution to the neutron
dripline. This includes the standard doubly-magic $^{48}$Ca at $N=28$
and new shell closures at $N=32$ and possibly at $N=34$. The calcium
chain, with only valence neutrons, probes features of nuclear forces
similar to the oxygen isotopes, which have been under intensive
experimental and theoretical studies~\cite{Thoenessen2012,Otsuka2010}.
While the oxygen isotopes have been explored even beyond the neutron
dripline, $^{52}$Ca is the most neutron-rich nucleus where mass 
measurements and $\gamma$-ray spectroscopy have been done. 
Due to the extremely low production yields for $N>32$,
only the neutron-rich, mid-shell titanium and chromium isotopes 
have been reached to $N=34$. This makes the semi-magic calcium chain
in the vicinity of $N=32$ an important stepping stone towards the neutron dripline.

Phenomenological forces in the $pf$ shell, such as the
KB3G~\cite{Poves2001} and GXPF1~\cite{Honma2002} interactions, have
been fit to $N \leqslant 32$ in the calcium isotopes, but they
disagree markedly in their prediction for $^{54}$Ca and for a possible
shell closure at $N=34$. In addition, it is well known that
calculations based only on two-nucleon (NN) forces do not reproduce
$^{48}$Ca as a doubly-magic nucleus when neutrons are added to a
$^{40}$Ca core~\cite{Caurier2005}. Motivated by these deficiencies,
the impact of 3N forces was recently investigated in the oxygen and
calcium isotopes within the shell model~\cite{Otsuka2010,%
Holt2010,Holt2011}. It was shown that chiral 3N forces provide
repulsive contributions to valence neutron-neutron interactions.
As neutrons are added, these are key for shell structure and 
spectroscopy, and for the determination of the neutron dripline. For
calcium, the $N=28$ magic number was reproduced successfully, with a
high $2^{+}$ excitation energy, and in the vicinity of $N=32$, 
the calculations based on NN and 3N forces generally predicted 
an increase in the binding energy compared to experimental 
masses~\cite{Holt2010}.

Nuclear masses provide important information about the interplay of
strong interactions within a nucleus and the resulting nuclear
structure effects they elicit. The systematic study of masses has,
for example, led to the discovery of the new magic number $N=16$ and has
been key for understanding the island of
inversion~\cite{Thibault1975,Sarazin2000,Jurado2007}. Currently,
Penning traps~\cite{Blaum2006} provide the most precise mass
spectrometers for stable and unstable nuclei and are in use at almost
all rare isotope beam facilities~\cite{Kluge2010}.

In this Letter, we present the first precision mass measurements of
$^{51,52}$Ca and $^{51}$K performed at TRIUMF's Ion Trap for
Atomic and Nuclear Science (TITAN)~\cite{Dilling2003,Dilling2006,%
Brodeur2012a}. This presents the first direct mass measurements to
$N = 32$ in these nuclides, extending the measurements of a previous
campaign~\cite{Lapierre2012}. For the calcium isotopes, we compare the
evolution of the new TITAN mass values with neutron number to microscopic
calculations based on chiral NN and 3N forces.

The $^{51,52}$Ca isotopes were produced at TRIUMF's Isotope Separator
and ACcelerator (ISAC) facility by bombarding a high-power tantalum
target with a 70~$\mu$A proton beam of 480~MeV energy, and a
resonant laser ionization scheme~\cite{TRILIS} was used to enhance the
ionization of the beam. $^{51}$K was produced with 1.9~$\mu$A of protons
on a UC$_{\textnormal{x}}$ target, and ionized using a surface
source. The continuous ion beam was extracted from the target, mass
separated with the ISAC dipole separator system, and delivered to the
TITAN facility. TITAN uses the time-of-flight ion cyclotron-resonance
(TOF-ICR)~\cite{Konig1995} method on singly-charged~\cite{Smith2008,%
Brodeur2012b,Ringle2009} and 
highly-charged~\cite{Ettenauer2011} ions for precision mass measurements and
is capable of accurate measurements in the parts-per-billion (ppb)
precision range~\cite{Brodeur2009}. The system consists of a
helium-buffer-gas-filled radio-frequency quadrupole~\cite{Brunner2012}
for cooling and bunching, followed by,
in this case bypassed, an electron beam ion trap charge breeder
(EBIT)~\cite{Lapierre2010}, and a precision
Penning trap~\cite{Brodeur2012a}, where the mass was determined.

\begin{figure}
\begin{center}
\includegraphics[width=0.48\textwidth,clip=]{%
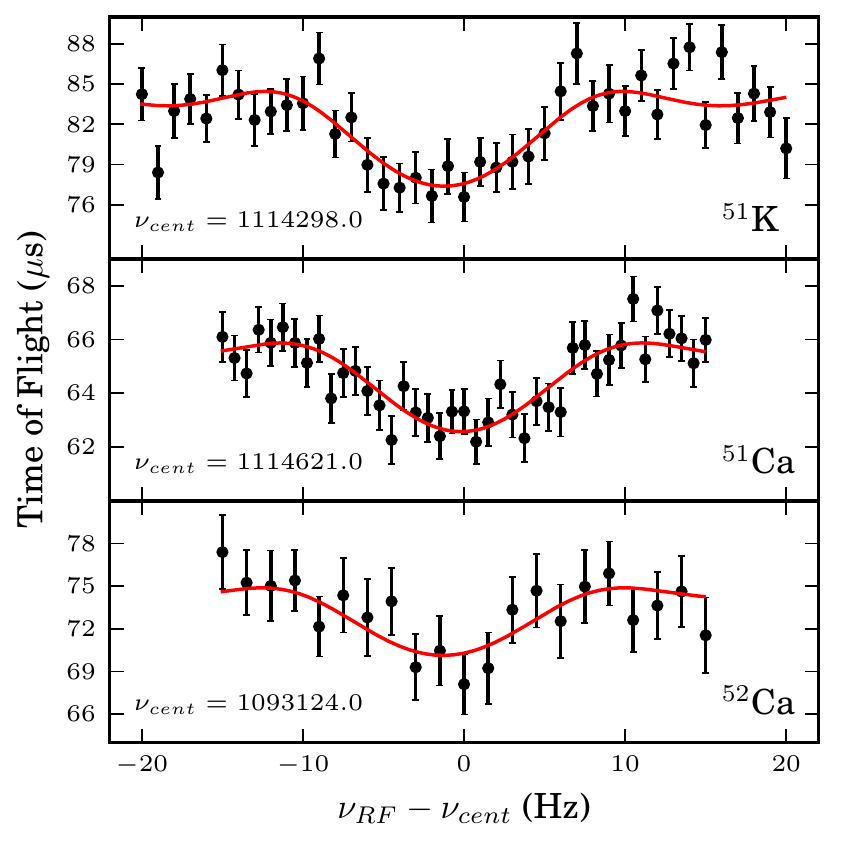}
\end{center}
\caption{(Color online) TOF-ICR resonances for $^{51}$K, $^{51}$Ca, 
and $^{52}$Ca with excitation times of $\approx80$~ms. The applied 
radio-frequency $\nu_{RF}$ is given as the difference from the 
center frequency $\nu_{cent}$ of the scan range.
The solid lines are theoretical fits to the line shape~\cite{Konig1995}.
\label{fig:TOFresonance_K51}}
\end{figure}

\begin{table*}[ht]
\caption{Measured average frequency ratios $\bar{R}$ and mass excess (ME) 
with total error of $^{51,52}$Ca and $^{51}$K,
compared to the AME2003 mass excess. The
* indicates that the AME2003 value for $^{51}$K is based on an 
extrapolation.\label{tab:results}}
\begin{tabular}{c c . d d ,}
\hline
\hline
Isotope & Reference & \multicolumn{1}{c}{$\bar{R}=\nu_{c,ref}/\nu_{c}$} 
& \multicolumn{1}{c}{ME~(keV)} & \multicolumn{1}{c}{ME$_{\rm AME03}$~(keV)} 
& \multicolumn{1}{c}{ME$-$ME$_{\rm AME03}$~(keV)} \\
\hline
$^{51}$Ca & $^{58}$Ni 
& 0.87961718(42) & -36338.9(22.7) & -35863.3(93.8) & -475.6(96.5) \\
\hline
$^{52}$Ca & $^{58}$Ni 
& 0.89691649(187) & -34260.0(101.0) & -32509.1(698.6) & -1751.0 \\
$^{52}$Ca & $^{52}$Cr 
& 1.00043782(158) & -34235.8(76.4) & {\rm same} & -1726.7 \\
\small{$^{52}$Ca Average:} & 
& & -34244.6(61.0) & {\rm same} & -1735.5(701.3) \\
\hline
$^{51}$K & $^{51}$V 
& 1.00062561(28) & -22516.3(13.1) & -22002.0(503.0)* & -514.3(503.2) \\ 
\hline
\hline
\end{tabular}
\end{table*}

In the precision Penning trap a homogenous 3.7~T magnetic field radially confines the
injected ions, while an electric quadrupole field provides axial
confinement. In order to determine an ion's mass $m$, the cyclotron
frequency $\nu_{c} = q B / (2 \pi m)$, where $q$ is the charge of the
ion and $B$ is a homogeneous magnetic field, is determined from the
minimum of the TOF-ICR measurement.
Typical TOF-ICR resonances for $^{51}$K,
$^{51}$Ca, and $^{52}$Ca are shown in Fig.~\ref{fig:TOFresonance_K51}. 
To calibrate the magnetic field, measurements with a reference ion of
well-known mass were taken before and after the frequency measurement
of the ion of interest. To eliminate magnetic field fluctuations a
linear interpolation of the reference frequency to the center time of
the measurement of $\nu_{c}$ is performed, and a ratio of the
frequencies $R = \nu_{c,ref} / \nu_{c}$ is taken. The atomic mass
$m_{a}$ of interest can then be extracted from the average frequency
ratio $\bar{R}$, $m_{a} = \bar{R} (m_{a,ref} - m_{e}) + m_{e}$ where
$m_{e}$ is the mass of the electron.

To reduce systematic effects, such as those arising from field
misalignments, incomplete trap compensation, etc., a reference with a
similar mass was chosen~\cite{Brodeur2012a}. For the measurements
where a mass doublet was not formed a mass measurement of $^{41}$K
relative to $^{58}$Ni was performed to investigate the mass-dependent
shift. The mass of $^{41}$K was in agreement with Ref.~\cite{Mount2010},
constraining the mass-dependent shift to be below 3.5 ppb in the
frequency ratio. We include this shift as a systematic
uncertainty. Moreover, to exclude potential effects stemming from
ion-ion interactions from simultaneously stored isobars, contaminants
were eliminated from the trap by applying a dipolar field at the mass-dependent
reduced cyclotron frequency $\nu_{+}$ for 20~ms prior to the
quadrupole excitation. Dipole cleaning pushes the
contaminants far from the precision confinement volume, greatly
reducing any potential shifts due to their Coulomb interaction with the ion
of interest. A quadrupole excitation time of $\approx 80$~ms
was used for each of the mass measurements.
In addition, a count class analysis~\cite{Kellerbauer2003} was performed
when the count rate was high enough to permit such an analysis. In
cases where the rate was too low, the frequency ratio was determined
twice: Once where only one ion was detected, and a second time where
all detected ions were included. The ion-ion interaction systematic
uncertainty was taken to be the difference between these two methods
and was 100~ppb for $^{51}$K and 750~ppb for $^{52}$Ca.

\begin{figure}[t]
\begin{center}
\includegraphics[width=0.48\textwidth,clip=]{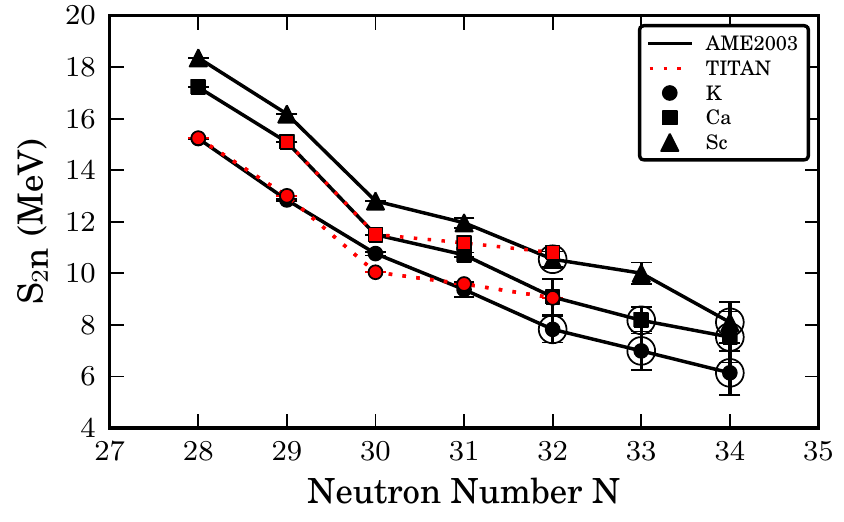}
\end{center}
\caption{(Color online) Two-neutron separation energy $S_{2n}$ as a
function of neutron number $N$ for the potassium (circles), calcium
(square), and scandium (triangles) isotopic chains. Points enclosed 
by a circle
are not based completely on experimental values in AME2003. The symbols
connected by a dotted line are based on the TITAN mass values (mass values
for $^{49,50}$Ca and $^{47-50}$K are taken from Ref.~\cite{Lapierre2012}),
while the symbols connected by solid lines are those from 
the AME2003.\label{fig:S2n}}
\end{figure}

In previous works, mass measurements for neutron-rich potassium and
calcium isotopes were reported up to $^{50}$K and
$^{52}$Ca~\cite{Tu1990,Audi2003}, but with large uncertainties. While
the half-lives in the vicinity of $N=32$ are generally long ($t_{1/2}
> 50$~ms), the production yields in this high $N/Z$ region have
limited precision mass measurements until now. The mass of $^{51}$Ca,
as tabulated in the atomic mass evaluation (AME2003)~\cite{Audi2003},
depends largely on three-neutron-transfer reactions from beams of
$^{14}$C or $^{18}$O to a $^{48}$Ca target~\cite{Mayer1980,Benenson1985,Brauner1985,Catford1988}.
Measurements bydifferent groups
using the same reaction lead to differing results, calling into
question the derived mass. Further problems with the reaction methods
include potential $^{40}$Ca contamination in the target, disagreement
on the number and excitation energies of observed states, and low
statistics. In addition, two TOF measurements of the
$^{51}$Ca mass~\cite{Tu1990,Seifert1994} are in agreement with each
other, but disagree with the values derived from the
three-neutron-transfer reactions. However, the uncertainties reported
are $2$--$10$ times larger than those obtained from the reactions. All
masses tabulated in the AME2003 disagree with the presented
measurement by more than 1$\sigma$. More recently, a TOF mass
measurement was completed at the GSI storage ring~\cite{Matos}. This
does not agree with any of the previous measurements and deviates by
1.3$\sigma$ from the value presented in this Letter.  For $^{52}$Ca,
the existing mass value is derived from a TOF measurement~\cite{Tu1990} and a
$\beta$-decay measurement to $^{52}$Sc~\cite{Huck1985}. Neither
measurement agrees with our precision mass. No experimental data
exists for the mass of $^{51}$K.

Our new TITAN mass measurements for $^{51,52}$Ca and $^{51}$K are
presented in Table~\ref{tab:results}. The mass of $^{51}$Ca deviates
by 5$\sigma$ from the AME2003 and is more bound by 0.5~MeV. We find a
similar increase in binding for $^{51}$K compared to the AME2003
extrapolation. For the most neutron-rich $^{52}$Ca isotope measured,
the mass is more bound by 1.74~MeV compared to the present mass table.
This dramatic increase in binding leads to a pronounced change of the
derived two-neutron separation energy $S_{2n}$ in the vicinity of $N=32$, as
shown in Fig.~\ref{fig:S2n}. The resulting behavior of $S_{2n}$ in the
potassium and calcium isotopic chains with increasing neutron number
is significantly flatter from $N=30$ to $N=32$. This also differs from the
scandium isotopes, derived from previously measured mass excess with
large uncertainties or from the AME2003 extrapolation. The increased
binding for the potassium and calcium isotopes may indicate the development
of a significant subshell gap at $N=32$, in line with the observed high 
$2^+$ excitation energy in $^{52}$Ca\cite{Huck1985}.

Three-nucleon forces have been unambiguously established in light
nuclei, but only recently explored in medium-mass
nuclei~\cite{Otsuka2010,Holt2010,Holt2011,Roth2011,Hagen2012}. These
advances have been driven by chiral effective field theory, which
provides a systematic expansion for NN, 3N and higher-body
forces~\cite{Epelbaum2008}, combined with renormalization group
methods to evolve nuclear forces to lower
resolution~\cite{Bogner2009}.

\begin{figure}[t]
\begin{center}
\includegraphics[width=0.38\textwidth,clip=]{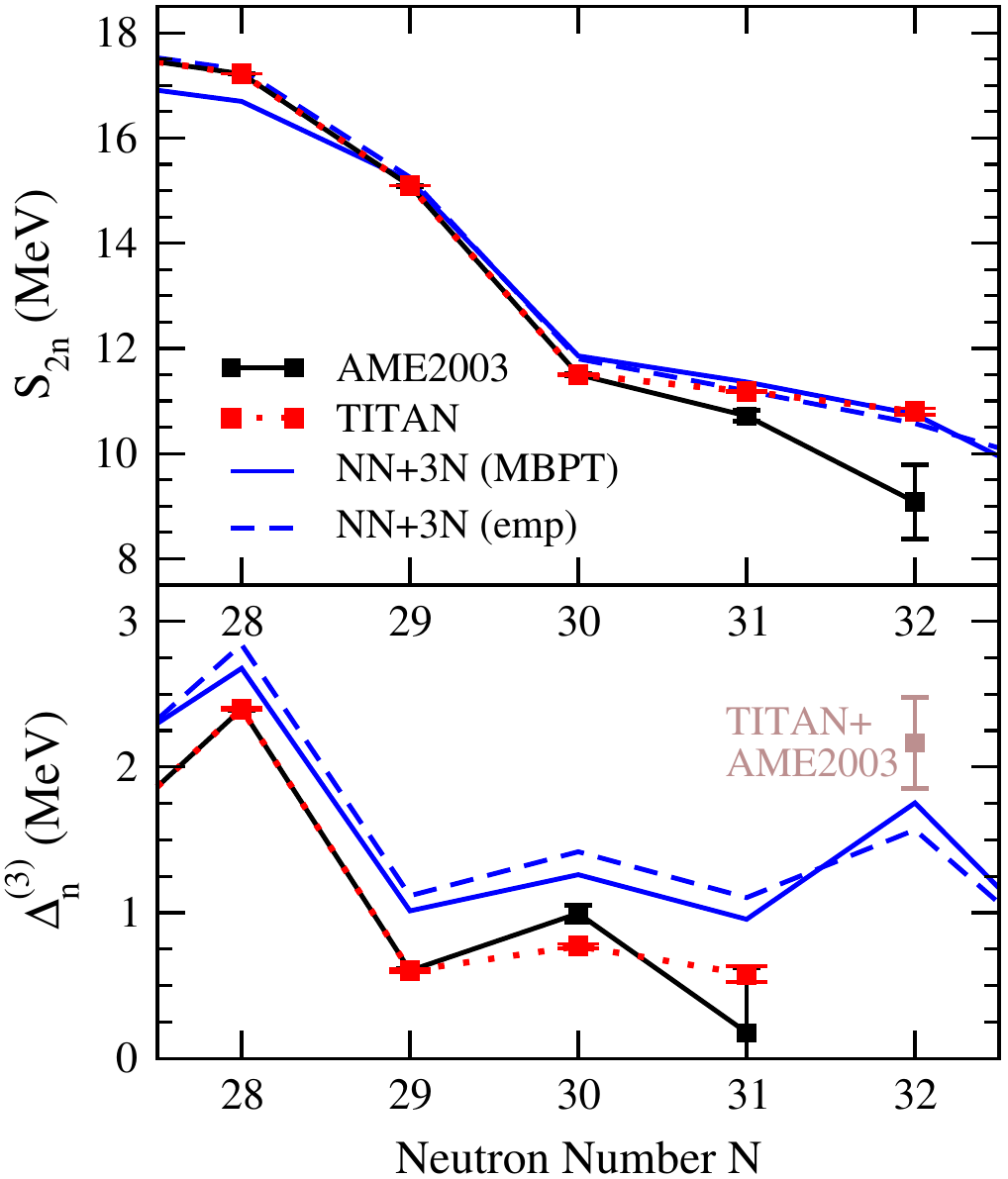}
\end{center}
\caption{(Color online) Two-neutron separation energy $S_{2n}$
(top) and pairing gap $\Delta_n^{(3)}$ from
three-point binding-energy
differences (bottom)  versus neutron number $N$ for the calcium isotopes.
The TITAN mass values and the AME2003 values are shown by the symbols
as in Fig.~\ref{fig:S2n}. The point labelled ``TITAN+AME2003'' is based partly on the TITAN mass values and complemented by the AME2003 value for $^{53}$Ca.
Theoretical predictions are shown based on chiral
NN and 3N forces (NN+3N) in the extended $pfg_{9/2}$ valence space
using empirical (emp) SPEs in $^{41}$Ca and consistently calculated
MBPT SPEs (MBPT).\label{fig:theory}}
\end{figure}

We follow Ref.~\cite{Holt2010} and calculate the two-body interactions
among valence neutrons in the extended $pfg_{9/2}$ shell on top of a
$^{40}$Ca core, taking into account valence-core interactions in 13
major shells based on a chiral N$^3$LO NN potential evolved to
low-momentum. Chiral 3N forces are included at N$^2$LO, where the two
shorter-range 3N couplings have been fit to the $^3$H binding energy and
the $^4$He charge radius. For valence neutrons the dominant
contribution is due to the long-range two-pion-exchange part of 3N
forces~\cite{Otsuka2010,Holt2010}. In Ref.~\cite{Holt2010}, the
normal-ordered one- and two-body parts of 3N forces were included to
first order, and an increased binding in the vicinity of $N=32$ was
predicted. Here, we improve the calculation by including the one- and
two-body parts of 3N forces in 5 major shells to third order, on an
equal footing as NN interactions. This takes into account the effect
of 3N forces between nucleons occupying orbits above and below the
valence space. For the single-particle energies (SPEs) in $^{41}$Ca,
we study two cases: The SPEs obtained by solving the Dyson equation,
where the self-energy is calculated consistently in many-body
perturbation theory (MBPT) to third order; and empirical (emp) SPEs
where the $pf$-orbit energies are taken from Ref.~\cite{Honma2002} and
the $g_{9/2}$ energy is set to $-1.0$~MeV.

In Fig.~\ref{fig:theory} we compare the theoretical results
obtained from exact diagonalizations in the valence space to the TITAN
and AME2003 values for the two-neutron separation energy and for the
neutron pairing gap calculated from three-point binding-energy differences,
$\Delta_n^{(3)}=(-1)^N [B(N+1,Z) +B(N-1,Z)-2B(N,Z)]/2$. The predicted
$S_{2n}$ is very similar for both sets of SPEs and is in excellent
agreement with the new TITAN mass values. For $^{51,52}$Ca, the difference
between theory and experiment is only $\lesssim 200$~keV, but we
emphasize that it will be important to also study the impact of the
uncertainties in the leading 3N forces. The behavior with neutron number for
$\Delta_n^{(3)}$ is also well reproduced, but the theoretical gaps are
typically 500~keV larger. Finally, we note also the developments using
nonempirical pairing functionals in this region~\cite{Lesinski2011},
which provide a bridge to global energy-density functional calculations.

In summary, the mass of $^{51}$K has been measured with the TITAN
facility at TRIUMF for the first time, and the new precision masses of
$^{51,52}$Ca show a dramatic increase in binding compared to the
atomic mass evaluation. The most neutron-rich $^{52}$Ca is more bound
by 1.74~MeV, a value similar in magnitude to the deuteron binding energy. An
increased binding around $N=32$ was predicted recently in calculations
based on chiral NN and 3N forces~\cite{Holt2010}. The new TITAN
results lead to a substantial change in the evolution of nuclear
masses to neutron-rich extremes. The significantly flatter behavior of
the two-neutron separation energy agrees remarkably well with improved
theoretical calculations including 3N forces, making neutron-rich
calcium isotopes an exciting region to probe 3N forces and to test
their predictions towards the neutron dripline. These developments are
of great interest also for astrophysics, as similar changes in heavier
nuclei would have a dramatic impact on nucleosynthesis~\cite{Arcones},
and the same 3N forces provide important repulsive contributions in
neutron-star matter~\cite{Hebeler2010}.

This work was supported by NSERC and the NRC Canada, the US DOE Grant
DE-FC02-07ER41457 (UNEDF SciDAC Collaboration) and DE-FG02-96ER40963
(UT), the Helmholtz Alliance HA216/EMMI, and the DFG through Grant SFB
634.  Part of the numerical calculations have been performed on Kraken
at the NICS. A.T.G.~acknowledges support from the NSERC CGS-D program,
S.E.~from the Vanier CGS program, T.B.~from the Evangelisches
Studienwerk e.V.~Villigst, V.V.S from the Studienstiftung des Deutschen Volkes,
and A.L.~from the DFG under Grant FR601/3-1.


\begin{thebibliography}{43}%
\makeatletter
\providecommand \@ifxundefined [1]{%
 \@ifx{#1\undefined}
}%
\providecommand \@ifnum [1]{%
 \ifnum #1\expandafter \@firstoftwo
 \else \expandafter \@secondoftwo
 \fi
}%
\providecommand \@ifx [1]{%
 \ifx #1\expandafter \@firstoftwo
 \else \expandafter \@secondoftwo
 \fi
}%
\providecommand \natexlab [1]{#1}%
\providecommand \enquote  [1]{``#1''}%
\providecommand \bibnamefont  [1]{#1}%
\providecommand \bibfnamefont [1]{#1}%
\providecommand \citenamefont [1]{#1}%
\providecommand \href@noop [0]{\@secondoftwo}%
\providecommand \href [0]{\begingroup \@sanitize@url \@href}%
\providecommand \@href[1]{\@@startlink{#1}\@@href}%
\providecommand \@@href[1]{\endgroup#1\@@endlink}%
\providecommand \@sanitize@url [0]{\catcode `\\12\catcode `\$12\catcode
  `\&12\catcode `\#12\catcode `\^12\catcode `\_12\catcode `\%12\relax}%
\providecommand \@@startlink[1]{}%
\providecommand \@@endlink[0]{}%
\providecommand \url  [0]{\begingroup\@sanitize@url \@url }%
\providecommand \@url [1]{\endgroup\@href {#1}{\urlprefix }}%
\providecommand \urlprefix  [0]{URL }%
\providecommand \Eprint [0]{\href }%
\providecommand \doibase [0]{http://dx.doi.org/}%
\providecommand \selectlanguage [0]{\@gobble}%
\providecommand \bibinfo  [0]{\@secondoftwo}%
\providecommand \bibfield  [0]{\@secondoftwo}%
\providecommand \translation [1]{[#1]}%
\providecommand \BibitemOpen [0]{}%
\providecommand \bibitemStop [0]{}%
\providecommand \bibitemNoStop [0]{.\EOS\space}%
\providecommand \EOS [0]{\spacefactor3000\relax}%
\providecommand \BibitemShut  [1]{\csname bibitem#1\endcsname}%
\let\auto@bib@innerbib\@empty
\bibitem [{\citenamefont {Baumann}\ \emph {et~al.}(2012)\citenamefont
  {Baumann}, \citenamefont {Spyrou},\ and\ \citenamefont
  {Thoennessen}}]{Thoenessen2012}%
  \BibitemOpen
  \bibfield  {author} {\bibinfo {author} {\bibfnamefont {T.}~\bibnamefont
  {Baumann}}, \bibinfo {author} {\bibfnamefont {A.}~\bibnamefont {Spyrou}}, \
  and\ \bibinfo {author} {\bibfnamefont {M.}~\bibnamefont {Thoennessen}},\
  }\href {\doibase 10.1088/0034-4885/75/3/036301} {\bibfield  {journal}
  {\bibinfo  {journal} {Rep. Prog. Phys.}\ }\textbf {\bibinfo {volume} {75}},\
  \bibinfo {pages} {036301} (\bibinfo {year} {2012})}\BibitemShut {NoStop}%
\bibitem [{\citenamefont {Otsuka~\textit{et al.}}(2010)}]{Otsuka2010}%
  \BibitemOpen
  \bibfield  {author} {\bibinfo {author} {\bibfnamefont {T.}~\bibnamefont
  {Otsuka~\textit{et al.}}},\ }\href {\doibase 10.1103/PhysRevLett.105.032501}
  {\bibfield  {journal} {\bibinfo  {journal} {Phys. Rev. Lett.}\ }\textbf
  {\bibinfo {volume} {105}},\ \bibinfo {pages} {032501} (\bibinfo {year}
  {2010})}\BibitemShut {NoStop}%
\bibitem [{\citenamefont {Poves~\textit{et al.}}(2001)}]{Poves2001}%
  \BibitemOpen
  \bibfield  {author} {\bibinfo {author} {\bibfnamefont {A.}~\bibnamefont
  {Poves~\textit{et al.}}},\ }\href {\doibase 10.1016/S0375-9474(01)00967-8}
  {\bibfield  {journal} {\bibinfo  {journal} {Nucl. Phys. A}\ }\textbf
  {\bibinfo {volume} {694}},\ \bibinfo {pages} {157} (\bibinfo {year}
  {2001})}\BibitemShut {NoStop}%
\bibitem [{\citenamefont {Honma~\textit{et al.}}(2002)}]{Honma2002}%
  \BibitemOpen
  \bibfield  {author} {\bibinfo {author} {\bibfnamefont {M.}~\bibnamefont
  {Honma~\textit{et al.}}},\ }\href {\doibase 10.1103/PhysRevC.65.061301}
  {\bibfield  {journal} {\bibinfo  {journal} {Phys. Rev. C}\ }\textbf {\bibinfo
  {volume} {65}},\ \bibinfo {pages} {061301(R)} (\bibinfo {year}
  {2002})}\BibitemShut {NoStop}%
\bibitem [{\citenamefont {Caurier~\textit{et al.}}(2005)}]{Caurier2005}%
  \BibitemOpen
  \bibfield  {author} {\bibinfo {author} {\bibfnamefont {E.}~\bibnamefont
  {Caurier~\textit{et al.}}},\ }\href {\doibase 10.1103/RevModPhys.77.427}
  {\bibfield  {journal} {\bibinfo  {journal} {Rev. Mod. Phys.}\ }\textbf
  {\bibinfo {volume} {77}},\ \bibinfo {pages} {427} (\bibinfo {year}
  {2005})}\BibitemShut {NoStop}%
\bibitem [{\citenamefont {Holt~\textit{et al.}}(2010)}]{Holt2010}%
  \BibitemOpen
  \bibfield  {author} {\bibinfo {author} {\bibfnamefont {J.~D.}\ \bibnamefont
  {Holt~\textit{et al.}}},\ }\href@noop {} {}\bibinfo {howpublished}
  {arXiv:1009.5984} (\bibinfo {year} {2010})\BibitemShut {NoStop}%
\bibitem [{\citenamefont {Holt~\textit{et al.}}(2011)}]{Holt2011}%
  \BibitemOpen
  \bibfield  {author} {\bibinfo {author} {\bibfnamefont {J.~D.}\ \bibnamefont
  {Holt~\textit{et al.}}},\ }\href@noop {} {}\bibinfo {howpublished}
  {arXiv:1108.2680} (\bibinfo {year} {2011})\BibitemShut {NoStop}%
\bibitem [{\citenamefont {Thibault~\textit{et al.}}(1975)}]{Thibault1975}%
  \BibitemOpen
  \bibfield  {author} {\bibinfo {author} {\bibfnamefont {C.}~\bibnamefont
  {Thibault~\textit{et al.}}},\ }\href {\doibase 10.1103/PhysRevC.12.644}
  {\bibfield  {journal} {\bibinfo  {journal} {Phys. Rev. C}\ }\textbf {\bibinfo
  {volume} {12}},\ \bibinfo {pages} {644} (\bibinfo {year} {1975})}\BibitemShut
  {NoStop}%
\bibitem [{\citenamefont {Sarazin~\textit{et al.}}(2000)}]{Sarazin2000}%
  \BibitemOpen
  \bibfield  {author} {\bibinfo {author} {\bibfnamefont {F.}~\bibnamefont
  {Sarazin~\textit{et al.}}},\ }\href {\doibase 10.1103/PhysRevLett.84.5062}
  {\bibfield  {journal} {\bibinfo  {journal} {Phys. Rev. Lett.}\ }\textbf
  {\bibinfo {volume} {84}},\ \bibinfo {pages} {5062} (\bibinfo {year}
  {2000})}\BibitemShut {NoStop}%
\bibitem [{\citenamefont {Jurado~\textit{et al.}}(2007)}]{Jurado2007}%
  \BibitemOpen
  \bibfield  {author} {\bibinfo {author} {\bibfnamefont {B.}~\bibnamefont
  {Jurado~\textit{et al.}}},\ }\href {\doibase 10.1016/j.physletb.2007.04.006}
  {\bibfield  {journal} {\bibinfo  {journal} {Phys. Lett. B}\ }\textbf
  {\bibinfo {volume} {649}},\ \bibinfo {pages} {43 } (\bibinfo {year}
  {2007})}\BibitemShut {NoStop}%
\bibitem [{\citenamefont {Blaum}(2006)}]{Blaum2006}%
  \BibitemOpen
  \bibfield  {author} {\bibinfo {author} {\bibfnamefont {K.}~\bibnamefont
  {Blaum}},\ }\href {\doibase 10.1016/j.physrep.2005.10.011} {\bibfield
  {journal} {\bibinfo  {journal} {Phys. Rep.}\ }\textbf {\bibinfo {volume}
  {425}},\ \bibinfo {pages} {1} (\bibinfo {year} {2006})}\BibitemShut {NoStop}%
\bibitem [{\citenamefont {Kluge}(2010)}]{Kluge2010}%
  \BibitemOpen
  \bibfield  {author} {\bibinfo {author} {\bibfnamefont {H.-J.}\ \bibnamefont
  {Kluge}},\ }\href@noop {} {\bibfield  {journal} {\bibinfo  {journal}
  {Hyperfine Interact.}\ }\textbf {\bibinfo {volume} {196}},\ \bibinfo {pages}
  {295} (\bibinfo {year} {2010})}\BibitemShut {NoStop}%
\bibitem [{\citenamefont {{Dilling \textit{et al.}}}(2003)}]{Dilling2003}%
  \BibitemOpen
  \bibfield  {author} {\bibinfo {author} {\bibfnamefont {J.}~\bibnamefont
  {{Dilling \textit{et al.}}}},\ }\href {\doibase
  10.1016/S0168-583X(02)02118-3} {\bibfield  {journal} {\bibinfo  {journal}
  {Nucl. Instrum. Methods Phys. Res. Sect. B}\ }\textbf {\bibinfo {volume}
  {204}},\ \bibinfo {pages} {492} (\bibinfo {year} {2003})}\BibitemShut
  {NoStop}%
\bibitem [{\citenamefont {Dilling~\textit{et al.}}(2006)}]{Dilling2006}%
  \BibitemOpen
  \bibfield  {author} {\bibinfo {author} {\bibfnamefont {J.}~\bibnamefont
  {Dilling~\textit{et al.}}},\ }\href {\doibase 10.1016/j.ijms.2006.01.044}
  {\bibfield  {journal} {\bibinfo  {journal} {Int. J. Mass Spectrom.}\ }\textbf
  {\bibinfo {volume} {251}},\ \bibinfo {pages} {198} (\bibinfo {year}
  {2006})}\BibitemShut {NoStop}%
\bibitem [{\citenamefont {{Brodeur \textit{et
  al.}}}(2012{\natexlab{a}})}]{Brodeur2012a}%
  \BibitemOpen
  \bibfield  {author} {\bibinfo {author} {\bibfnamefont {M.}~\bibnamefont
  {{Brodeur \textit{et al.}}}},\ }\href {\doibase 10.1016/j.ijms.2011.11.002}
  {\bibfield  {journal} {\bibinfo  {journal} {Int. J. Mass Spectrom.}\ }\textbf
  {\bibinfo {volume} {310}},\ \bibinfo {pages} {20} (\bibinfo {year}
  {2012}{\natexlab{a}})}\BibitemShut {NoStop}%
\bibitem [{\citenamefont {Lapierre~\textit{et
  al.}}(2012{\natexlab{a}})}]{Lapierre2012}%
  \BibitemOpen
  \bibfield  {author} {\bibinfo {author} {\bibfnamefont {A.}~\bibnamefont
  {Lapierre~\textit{et al.}}},\ }\href {\doibase 10.1103/PhysRevC.85.024317}
  {\bibfield  {journal} {\bibinfo  {journal} {Phys. Rev. C}\ }\textbf {\bibinfo
  {volume} {85}},\ \bibinfo {pages} {024317} (\bibinfo {year}
  {2012}{\natexlab{a}})}\BibitemShut {NoStop}%
\bibitem [{\citenamefont {{Lassen \textit{et al.}}}(2006)}]{TRILIS}%
  \BibitemOpen
  \bibfield  {author} {\bibinfo {author} {\bibfnamefont {J.}~\bibnamefont
  {{Lassen \textit{et al.}}}},\ }in\ \href {\doibase 10.1007/3-540-30926-8_8}
  {\emph {\bibinfo {booktitle} {Laser 2004}}},\ \bibinfo {editor} {edited by\
  \bibinfo {editor} {\bibfnamefont {Z.}~\bibnamefont {Błaszczak}}, \bibinfo
  {editor} {\bibfnamefont {B.}~\bibnamefont {Markov}}, \ and\ \bibinfo {editor}
  {\bibfnamefont {K.}~\bibnamefont {Marinova}}}\ (\bibinfo  {publisher}
  {Springer},\ \bibinfo {year} {2006})\ pp.\ \bibinfo {pages}
  {69--75}\BibitemShut {NoStop}%
\bibitem [{\citenamefont {K\"{o}nig~\textit{et al.}}(1995)}]{Konig1995}%
  \BibitemOpen
  \bibfield  {author} {\bibinfo {author} {\bibfnamefont {M.}~\bibnamefont
  {K\"{o}nig~\textit{et al.}}},\ }\href {\doibase 10.1016/0168-1176(95)04146-C}
  {\bibfield  {journal} {\bibinfo  {journal} {Int. J. Mass Spectrom.}\ }\textbf
  {\bibinfo {volume} {142}},\ \bibinfo {pages} {95} (\bibinfo {year}
  {1995})}\BibitemShut {NoStop}%
\bibitem [{\citenamefont {{Smith \textit{et al.}}}(2008)}]{Smith2008}%
  \BibitemOpen
  \bibfield  {author} {\bibinfo {author} {\bibfnamefont {M.}~\bibnamefont
  {{Smith \textit{et al.}}}},\ }\href {\doibase 10.1103/PhysRevLett.101.202501}
  {\bibfield  {journal} {\bibinfo  {journal} {Phys. Rev. Lett.}\ }\textbf
  {\bibinfo {volume} {101}},\ \bibinfo {pages} {202501} (\bibinfo {year}
  {2008})}\BibitemShut {NoStop}%
\bibitem [{\citenamefont {{Brodeur \textit{et
  al.}}}(2012{\natexlab{b}})}]{Brodeur2012b}%
  \BibitemOpen
  \bibfield  {author} {\bibinfo {author} {\bibfnamefont {M.}~\bibnamefont
  {{Brodeur \textit{et al.}}}},\ }\href {\doibase
  10.1103/PhysRevLett.108.052504} {\bibfield  {journal} {\bibinfo  {journal}
  {Phys. Rev. Lett.}\ }\textbf {\bibinfo {volume} {108}},\ \bibinfo {pages}
  {052504} (\bibinfo {year} {2012}{\natexlab{b}})}\BibitemShut {NoStop}%
\bibitem [{\citenamefont {{Ringle \textit{et al.}}}(2009)}]{Ringle2009}%
  \BibitemOpen
  \bibfield  {author} {\bibinfo {author} {\bibfnamefont {R.}~\bibnamefont
  {{Ringle \textit{et al.}}}},\ }\href {\doibase
  10.1016/j.physletb.2009.04.014} {\bibfield  {journal} {\bibinfo  {journal}
  {Phys. Lett. B}\ }\textbf {\bibinfo {volume} {675}},\ \bibinfo {pages} {170}
  (\bibinfo {year} {2009})}\BibitemShut {NoStop}%
\bibitem [{\citenamefont {{Ettenauer \textit{et al.}}}(2011)}]{Ettenauer2011}%
  \BibitemOpen
  \bibfield  {author} {\bibinfo {author} {\bibfnamefont {S.}~\bibnamefont
  {{Ettenauer \textit{et al.}}}},\ }\href {\doibase
  10.1103/PhysRevLett.107.272501} {\bibfield  {journal} {\bibinfo  {journal}
  {Phys. Rev. Lett.}\ }\textbf {\bibinfo {volume} {107}},\ \bibinfo {pages}
  {212501} (\bibinfo {year} {2011})}\BibitemShut {NoStop}%
\bibitem [{\citenamefont {{Brodeur \textit{et al.}}}(2009)}]{Brodeur2009}%
  \BibitemOpen
  \bibfield  {author} {\bibinfo {author} {\bibfnamefont {M.}~\bibnamefont
  {{Brodeur \textit{et al.}}}},\ }\href {\doibase 10.1103/PhysRevC.80.044318}
  {\bibfield  {journal} {\bibinfo  {journal} {Phys. Rev. C}\ }\textbf {\bibinfo
  {volume} {80}},\ \bibinfo {pages} {044318} (\bibinfo {year}
  {2009})}\BibitemShut {NoStop}%
\bibitem [{\citenamefont {Brunner~\textit{et al.}}(2012)}]{Brunner2012}%
  \BibitemOpen
  \bibfield  {author} {\bibinfo {author} {\bibfnamefont {T.}~\bibnamefont
  {Brunner~\textit{et al.}}},\ }\href {\doibase 10.1016/j.nima.2012.02.004}
  {\bibfield  {journal} {\bibinfo  {journal} {Nucl. Instrum. Methods Phys. Res.
  Sect. A}\ }\textbf {\bibinfo {volume} {676}},\ \bibinfo {pages} {32 }
  (\bibinfo {year} {2012})}\BibitemShut {NoStop}%
\bibitem [{\citenamefont {Lapierre~\textit{et
  al.}}(2012{\natexlab{b}})}]{Lapierre2010}%
  \BibitemOpen
  \bibfield  {author} {\bibinfo {author} {\bibfnamefont {A.}~\bibnamefont
  {Lapierre~\textit{et al.}}},\ }\href {\doibase 10.1016/j.nima.2010.09.030}
  {\bibfield  {journal} {\bibinfo  {journal} {Nucl. Instrum. Methods Phys. Res.
  Sect. A}\ }\textbf {\bibinfo {volume} {624}},\ \bibinfo {pages} {54}
  (\bibinfo {year} {2012}{\natexlab{b}})}\BibitemShut {NoStop}%
\bibitem [{\citenamefont {Mount}\ \emph {et~al.}(2010)\citenamefont {Mount},
  \citenamefont {Redshaw},\ and\ \citenamefont {Myers}}]{Mount2010}%
  \BibitemOpen
  \bibfield  {author} {\bibinfo {author} {\bibfnamefont {B.~J.}\ \bibnamefont
  {Mount}}, \bibinfo {author} {\bibfnamefont {M.}~\bibnamefont {Redshaw}}, \
  and\ \bibinfo {author} {\bibfnamefont {E.~G.}\ \bibnamefont {Myers}},\ }\href
  {\doibase 10.1103/PhysRevA.82.042513} {\bibfield  {journal} {\bibinfo
  {journal} {Phys. Rev. A}\ }\textbf {\bibinfo {volume} {82}},\ \bibinfo
  {pages} {042513} (\bibinfo {year} {2010})}\BibitemShut {NoStop}%
\bibitem [{\citenamefont {Kellerbauer~\textit{et
  al.}}(2003)}]{Kellerbauer2003}%
  \BibitemOpen
  \bibfield  {author} {\bibinfo {author} {\bibfnamefont {A.}~\bibnamefont
  {Kellerbauer~\textit{et al.}}},\ }\href {\doibase 10.1140/epjd/e2002-00222-0}
  {\bibfield  {journal} {\bibinfo  {journal} {Eur. Phys. J. D}\ }\textbf
  {\bibinfo {volume} {22}},\ \bibinfo {pages} {53} (\bibinfo {year}
  {2003})}\BibitemShut {NoStop}%
\bibitem [{\citenamefont {Tu~\textit{et al.}}(1990)}]{Tu1990}%
  \BibitemOpen
  \bibfield  {author} {\bibinfo {author} {\bibfnamefont {X.~L.}\ \bibnamefont
  {Tu~\textit{et al.}}},\ }\href {\doibase 10.1007/BF01294971} {\bibfield
  {journal} {\bibinfo  {journal} {Z. Phys. A: At. Nucl.}\ }\textbf {\bibinfo
  {volume} {337}},\ \bibinfo {pages} {361} (\bibinfo {year}
  {1990})}\BibitemShut {NoStop}%
\bibitem [{\citenamefont {Audi}\ \emph {et~al.}(2003)\citenamefont {Audi},
  \citenamefont {Wapstra},\ and\ \citenamefont {Thibault}}]{Audi2003}%
  \BibitemOpen
  \bibfield  {author} {\bibinfo {author} {\bibfnamefont {G.}~\bibnamefont
  {Audi}}, \bibinfo {author} {\bibfnamefont {A.~H.}\ \bibnamefont {Wapstra}}, \
  and\ \bibinfo {author} {\bibfnamefont {C.}~\bibnamefont {Thibault}},\ }\href
  {\doibase 10.1016/j.nuclphysa.2003.11.003} {\bibfield  {journal} {\bibinfo
  {journal} {Nucl. Phys. A}\ }\textbf {\bibinfo {volume} {729}},\ \bibinfo
  {pages} {337} (\bibinfo {year} {2003})}\BibitemShut {NoStop}%
\bibitem [{\citenamefont {Mayer~\textit{et al.}}(1980)}]{Mayer1980}%
  \BibitemOpen
  \bibfield  {author} {\bibinfo {author} {\bibfnamefont {W.}~\bibnamefont
  {Mayer~\textit{et al.}}},\ }\href {\doibase 10.1103/PhysRevC.22.2449}
  {\bibfield  {journal} {\bibinfo  {journal} {Phys. Rev. C}\ }\textbf {\bibinfo
  {volume} {22}},\ \bibinfo {pages} {2449} (\bibinfo {year}
  {1980})}\BibitemShut {NoStop}%
\bibitem [{\citenamefont {Benenson~\textit{et al.}}(1985)}]{Benenson1985}%
  \BibitemOpen
  \bibfield  {author} {\bibinfo {author} {\bibfnamefont {W.}~\bibnamefont
  {Benenson~\textit{et al.}}},\ }\href {\doibase 10.1016/0370-2693(85)91066-4}
  {\bibfield  {journal} {\bibinfo  {journal} {Phys. Lett. B}\ }\textbf
  {\bibinfo {volume} {162}},\ \bibinfo {pages} {87} (\bibinfo {year}
  {1985})}\BibitemShut {NoStop}%
\bibitem [{\citenamefont {Brauner~\textit{et al.}}(1985)}]{Brauner1985}%
  \BibitemOpen
  \bibfield  {author} {\bibinfo {author} {\bibfnamefont {M.}~\bibnamefont
  {Brauner~\textit{et al.}}},\ }\href {\doibase 10.1016/0370-2693(85)90141-8}
  {\bibfield  {journal} {\bibinfo  {journal} {Phys. Lett. B}\ }\textbf
  {\bibinfo {volume} {150}},\ \bibinfo {pages} {75} (\bibinfo {year}
  {1985})}\BibitemShut {NoStop}%
\bibitem [{\citenamefont {Catford~\textit{et al.}}(1988)}]{Catford1988}%
  \BibitemOpen
  \bibfield  {author} {\bibinfo {author} {\bibfnamefont {W.~N.}\ \bibnamefont
  {Catford~\textit{et al.}}},\ }\href {\doibase 10.1016/0375-9474(88)90157-1}
  {\bibfield  {journal} {\bibinfo  {journal} {Nucl. Phys. A}\ }\textbf
  {\bibinfo {volume} {489}},\ \bibinfo {pages} {347} (\bibinfo {year}
  {1988})}\BibitemShut {NoStop}%
\bibitem [{\citenamefont {Seifert~\textit{et al.}}(1994)}]{Seifert1994}%
  \BibitemOpen
  \bibfield  {author} {\bibinfo {author} {\bibfnamefont {H.~L.}\ \bibnamefont
  {Seifert~\textit{et al.}}},\ }\href {\doibase 10.1007/BF01296329} {\bibfield
  {journal} {\bibinfo  {journal} {Z. Phys. A: At. Nucl.}\ }\textbf {\bibinfo
  {volume} {349}},\ \bibinfo {pages} {25} (\bibinfo {year} {1994})}\BibitemShut
  {NoStop}%
\bibitem [{\citenamefont {Mato\v{s}}(2004)}]{Matos}%
  \BibitemOpen
  \bibfield  {author} {\bibinfo {author} {\bibfnamefont {M.}~\bibnamefont
  {Mato\v{s}}},\ }\href@noop {} {Ph.D. thesis},\ \bibinfo  {school}
  {Justus-Liebig-Universit\"{a}t Giessen} (\bibinfo {year} {2004})\BibitemShut
  {NoStop}%
\bibitem [{\citenamefont {Huck~\textit{et al.}}(1985)}]{Huck1985}%
  \BibitemOpen
  \bibfield  {author} {\bibinfo {author} {\bibfnamefont {A.}~\bibnamefont
  {Huck~\textit{et al.}}},\ }\href {\doibase 10.1103/PhysRevC.31.2226}
  {\bibfield  {journal} {\bibinfo  {journal} {Phys. Rev. C}\ }\textbf {\bibinfo
  {volume} {31}},\ \bibinfo {pages} {2226} (\bibinfo {year}
  {1985})}\BibitemShut {NoStop}%
\bibitem [{\citenamefont {Roth~\textit{et al.}}(2011)}]{Roth2011}%
  \BibitemOpen
  \bibfield  {author} {\bibinfo {author} {\bibfnamefont {R.}~\bibnamefont
  {Roth~\textit{et al.}}},\ }\href@noop {} {}\bibinfo {howpublished}
  {arXiv:1112.0287} (\bibinfo {year} {2011})\BibitemShut {NoStop}%
\bibitem [{\citenamefont {Hagen~\textit{et al.}}(2012)}]{Hagen2012}%
  \BibitemOpen
  \bibfield  {author} {\bibinfo {author} {\bibfnamefont {G.}~\bibnamefont
  {Hagen~\textit{et al.}}},\ }\href@noop {} {}\bibinfo {howpublished}
  {arXiv:1202.2839} (\bibinfo {year} {2012})\BibitemShut {NoStop}%
\bibitem [{\citenamefont {Epelbaum}\ \emph {et~al.}(2009)\citenamefont
  {Epelbaum}, \citenamefont {Hammer},\ and\ \citenamefont
  {Mei{\ss}ner}}]{Epelbaum2008}%
  \BibitemOpen
  \bibfield  {author} {\bibinfo {author} {\bibfnamefont {E.}~\bibnamefont
  {Epelbaum}}, \bibinfo {author} {\bibfnamefont {H.-W.}\ \bibnamefont
  {Hammer}}, \ and\ \bibinfo {author} {\bibfnamefont {U.-G.}\ \bibnamefont
  {Mei{\ss}ner}},\ }\href {\doibase 10.1103/RevModPhys.81.1773} {\bibfield
  {journal} {\bibinfo  {journal} {Rev. Mod. Phys.}\ }\textbf {\bibinfo {volume}
  {81}},\ \bibinfo {pages} {1773} (\bibinfo {year} {2009})}\BibitemShut
  {NoStop}%
\bibitem [{\citenamefont {Bogner}\ \emph {et~al.}(2010)\citenamefont {Bogner},
  \citenamefont {Furnstahl},\ and\ \citenamefont {Schwenk}}]{Bogner2009}%
  \BibitemOpen
  \bibfield  {author} {\bibinfo {author} {\bibfnamefont {S.~K.}\ \bibnamefont
  {Bogner}}, \bibinfo {author} {\bibfnamefont {R.~J.}\ \bibnamefont
  {Furnstahl}}, \ and\ \bibinfo {author} {\bibfnamefont {A.}~\bibnamefont
  {Schwenk}},\ }\href {\doibase 10.1016/j.ppnp.2010.03.001} {\bibfield
  {journal} {\bibinfo  {journal} {Prog. Part. Nucl. Phys.}\ }\textbf {\bibinfo
  {volume} {65}},\ \bibinfo {pages} {94} (\bibinfo {year} {2010})}\BibitemShut
  {NoStop}%
\bibitem [{\citenamefont {Lesinski~\textit{et al.}}(2012)}]{Lesinski2011}%
  \BibitemOpen
  \bibfield  {author} {\bibinfo {author} {\bibfnamefont {T.}~\bibnamefont
  {Lesinski~\textit{et al.}}},\ }\href {\doibase 10.1088/0954-3899/39/1/015108}
  {\bibfield  {journal} {\bibinfo  {journal} {J. Phys. G}\ }\textbf {\bibinfo
  {volume} {39}},\ \bibinfo {pages} {015108} (\bibinfo {year}
  {2012})}\BibitemShut {NoStop}%
\bibitem [{\citenamefont {Arcones}\ and\ \citenamefont
  {Bertsch}(2011)}]{Arcones}%
  \BibitemOpen
  \bibfield  {author} {\bibinfo {author} {\bibfnamefont {A.}~\bibnamefont
  {Arcones}}\ and\ \bibinfo {author} {\bibfnamefont {G.~F.}\ \bibnamefont
  {Bertsch}},\ }\href@noop {} {}\bibinfo {howpublished} {arXiv:1111.4923}
  (\bibinfo {year} {2011})\BibitemShut {NoStop}%
\bibitem [{\citenamefont {Hebeler~\textit{et al.}}(2010)}]{Hebeler2010}%
  \BibitemOpen
  \bibfield  {author} {\bibinfo {author} {\bibfnamefont {K.}~\bibnamefont
  {Hebeler~\textit{et al.}}},\ }\href {\doibase 10.1103/PhysRevLett.105.161102}
  {\bibfield  {journal} {\bibinfo  {journal} {Phys. Rev. Lett.}\ }\textbf
  {\bibinfo {volume} {105}},\ \bibinfo {pages} {161102} (\bibinfo {year}
  {2010})}\BibitemShut {NoStop}%
\end{thebibliography}
\end{document}